\pdfoutput=1
\documentclass[a4paper]{report}
\usepackage[utf8]{inputenc}
\usepackage[T1]{fontenc}
\usepackage{RJournal}
\usepackage{amsmath,amssymb,array}
\usepackage{booktabs}

\usepackage[labelformat=simple]{subcaption}

\usepackage{tikz,enumitem}
\usetikzlibrary{positioning, arrows.meta, shapes.geometric}
\tikzset{%
  semithick,
  >={Stealth[width=2mm, length=2.75mm]},
  obs/.style = {name = #1, circle, draw, inner sep = 5pt, label = center:\(\scriptstyle#1\)},
  fixed/.style = {name = #1, regular polygon, regular polygon sides = 4, draw, inner sep = 3pt, label = center:\(#1\)},
  lat/.style 2 args = {name = #1, circle, draw, dashed, inner sep = 5pt, label = center:\(\scriptstyle#2\)},
}
\newcommand\independent{\protect\mathpalette{\protect\independenT}{\perp}}
\def\independenT#1#2{\mathrel{\rlap{\(#1#2\)}\mkern2mu{#1#2}}}
\newcommand{\+}[1]{\ensuremath{\mathbf{#1}}}
\newcommand{\doo}{\textrm{do}}

\begin{document}

\sectionhead{ }
\volume{}
\volnumber{}
\year{}
\month{}

\begin{article}

  \title{Identifying Counterfactual Queries with the R Package \pkg{cfid}}
  \author{by Santtu Tikka}
  
  
  \maketitle
  
  \abstract{%
  In the framework of structural causal models, counterfactual queries describe events that concern multiple alternative states of the system under study. Counterfactual queries often take the form of ``what if'' type questions such as ``would an applicant have been hired if they had over 10 years of experience, when in reality they only had 5 years of experience?'' Such questions and counterfactual inference in general are crucial, for example when addressing the problem of fairness in decision-making. Because counterfactual events contain contradictory states of the world, it is impossible to conduct a randomized experiment to address them without making several restrictive assumptions. However, it is sometimes possible to identify such queries from observational and experimental data by representing the system under study as a causal model, and the available data as symbolic probability distributions. \citet{shpitser2007} constructed two algorithms, called ID* and IDC*, for identifying counterfactual queries and conditional counterfactual queries, respectively. These two algorithms are analogous to the ID and IDC algorithms by \citet{shpitser2006id, shpitser2006idc} for identification of interventional distributions, which were implemented in R by \citet{tikka2017} in the \pkg{causaleffect} package. We present the R package \CRANpkg{cfid} that implements the ID* and IDC* algorithms. Identification of counterfactual queries and the features of \pkg{cfid} are demonstrated via examples.
  }
  
  
  \section{Introduction} \label{sec:intro}
  
  Pearl's ladder of causation (or causal hierarchy) consists of three levels: association, intervention, and counterfactual \citep{pearl2009}. These levels describe a hierarchy of problems with increasing conceptual and formal difficulty. On the first and lowest level, inference on associations is based entirely on observed data in the form of questions such as ``what is the probability that an event occurs?'' or ``what is the correlation between two variables''. On the second level, the inference problems are related to manipulations of the system under study such as ``what is the probability of an event if we change the value of one variable in the system''. Questions on the intervention level cannot be answered using tools of the association level, because simply observing a change in a system is not the same as intervening on the system. Randomized controlled trials are the gold standard for studying the effects of interventions, because they enable the researcher to account for confounding factors between the treatment and the outcome and to carry out the intervention in practice. However, there are often practical limitations that make it difficult, expensive, or impossible to conduct a randomized experiment. The third and highest level is the counterfactual level. Typically, counterfactual statements compare the real world, where an action was taken or some event was observed, to an alternative hypothetical scenario, where a possibly different action was taken, or a different event was observed. Counterfactuals are often challenging to understand even conceptually due this notion of contradictory events in alternative worlds, and such alternatives need not be limited to only two. In general, questions on the counterfactual level cannot be answered by relying solely on the previous levels: no intervention or association is able to capture the notion of alternative hypothetical worlds.
  
  While counterfactual statements can be challenging, they are a core part of our everyday thinking and discourse. Importantly, counterfactuals often consider retrospective questions about the state of the world, such as ``would an applicant have been hired if they had more work experience''. This kind of retrospection is crucial when fair treatment of individuals is considered in hiring, healthcare, receiving loans or insurance, etc., with regards to protected attributes, especially when the goal is automated decision-making. Statistical approaches to fairness are insufficient in most contexts, such as in scenarios analogous to the well-known Simpson's paradox, but routinely resolved using the framework of causal inference. In some cases, even interventional notions of fairness may be insufficient, necessitating counterfactual fairness \citep{KusnerCounterfactual, ZhangBareinboim2018}.
  
  
  The structural causal model (SCM) framework of Pearl provides a formal approach to causal inference of interventional and counterfactual causal queries \citep{pearl2009}. An SCM represents the system of interest in two ways, First, the causal relationships are depicted by a directed acyclic graph (DAG) whose vertices correspond to variables under study and whose edges depict the direct functional causal relationships between the variables. Typically, only some of these variables are observed and the remaining variables are considered latent, corresponding either to confounders between multiple variables or individual random errors of single variables. Second, the uncertainty related to the variables in the system is captured by assuming a joint probability distribution over its latent variables. The functional relationships of the model induce a joint probability distribution over the observed variables. The SCM framework also incorporates the notion of external interventions symbolically via the do-operator, and a graphical representation of counterfactual scenarios via parallel worlds graphs \citep{avin2005pathspecific, shpitser2007, shpitser2008}.
  
  
  One of the fundamental problems of causal inference is the so-called identifiability problem, especially the identifiability of interventional distributions. Using the SCM framework and do-calculus, it is sometimes possible to uniquely represent an interventional distribution using only the observed joint probability distribution of the model before the intervention took place. Such interventional distributions are called \dfn{identifiable}. More generally, we say that a causal query is identifiable, if it can be uniquely represented using the available data. In most identifiability problems, the available data consists of causal quantities on levels below the query in the ladder of causation, but the levels also sometimes overlap, \citep[e.g.,][]{bareinboim2012zid,tikka2019surrogate,lee2019surrogate}. The identifiability problem of interventional distributions, and many other interventional identifiability problems have been solved by providing a sound and complete identification algorithm \citep[e.g.,][]{shpitser2006id, huang2006complete, lee2019surrogate, kivva2022}.
  
  Software for causal inference is becoming increasingly prominent. For R, a comprehensive overview of the state-of-the-art is provided by the recently launched task view on \ctv{Causal Inference} on the Comprehensive R Archive Network (CRAN). Out of the packages listed in this task view, the \CRANpkg{Counterfactual} \citep{counterfactualpackage} and \CRANpkg{WhatIf} \citep{whatifpackage} packages are directly linked to counterfactual inference, but the focus of these packages is estimation and they do not consider the identifiability of counterfactual queries. The \CRANpkg{R6causal} \citep{r6causal} package can be used to simulate data from counterfactual scenarios in a causal model. R packages most closely related to causal identifiability problems are the \CRANpkg{causaleffect} \citep{tikka2017}, \CRANpkg{dosearch} \citep{dosearch}, and \CRANpkg{dagitty} \citep{dagitty}.
  
  We present the first implementation of the counterfactual identifiability algorithms of \citet{shpitser2007} \citep[see also][]{shpitser2008} as the R package \pkg{cfid} (\textbf{c}ounter\textbf{f}actual \textbf{id}entification). The \pkg{cfid} package also provides a user-friendly interface for defining causal diagrams and the package is compatible with other major R packages for causal identifiability problems such as \pkg{causaleffect}, \pkg{dosearch} and \pkg{dagitty} by supporting graph formats used by these packages as inputs.
  
  The paper is organized as follows. Section~\ref{sec:notation} introduces the notation, core concepts and definitions, and provides an example on manual identification of a counterfactual query without relying on the identifiability algorithms. Section~\ref{sec:algo} presents the algorithms implemented in \pkg{cfid} and demonstrates their functionality via examples by tracing their operation line by line. Section~\ref{sec:package} demonstrates the usage of the \pkg{cfid} package in practice. Section~\ref{sec:summary} concludes the paper with a summary.

  \section{Notation and definitions} \label{sec:notation}
  
  We follow the notation used by \citet{shpitser2008} and we assume the reader to be familiar with standard graph theoretic concepts such as ancestral relations between vertices and d-separation. We use capital letters to denote random variables and lower-case letters to denote their value assignments. Bold letters are used to denote sets of random variables and counterfactual variables. We associate the vertices of graphs with their respective random variables and value assignments in the underlying causal models. In figures, observed variables of graphs are denoted by circles, variables fixed by interventions are denoted by squares, and latent unobserved variables are denoted by dashed circles when explicitly included and by bidirected edges when the corresponding latent variable has two observed children. Latent variables with only one child, which are called \dfn{error terms}, are not shown for clarity. 
  
  A \dfn{structural causal model} is a tuple \(M = (\+ U, \+ V, \+ F, P(\+ u))\) where \(\+ U\) is a set of unobserved random variables, \(\+ V\) is a set of \(n\) observed random variables, \(\+ F\) is a set of \(n\) functions such that each function \(f_i\) is a mapping from \(\+ U \cup \+ V \setminus \{V_i\}\) to \(V_i\) and such that it is possible to represent the set \(\+ V\) as function of \(\+ U\). \(P(\+ u)\) is a joint probability distribution over \(\+ U\). The causal model also defines its causal diagram \(G\). Each \(V_i \in \+ V\) corresponds to a vertex in \(G\), and there is a directed edge from each \(V_j \in \+ U \cup \+ V \setminus \{V_i\}\) to \(V_i\). We restrict our attention to \dfn{recursive} causal models in this paper, meaning models that induce an acyclic causal diagram.
  
  A \dfn{counterfactual variable} \(Y_{\+ x}\) denotes the variable \(Y\) in the submodel \(M_{\+ x}\) obtained from \(M\) by forcing the random variables \(\+ X\) to take the values \(\+ x\) (often denoted by the do-operator as \(\doo(\+ X = \+ x)\) or simply \(\doo(\+ x)\)). The distribution of \(\+ Y_{\+ x}\) in the submodel \(M_{\+ x}\) is called the \dfn{interventional distribution} of \(\+ Y\) and it is denoted by \(P_{\+ x}(\+ y)\). However, if we wish to consider multiple counterfactual variables that originate from different interventions, we must extend our notation to counterfactual conjunctions. \dfn{Counterfactual conjunctions} are constructed from value assignments of counterfactual variables, and individual assignments are separated by the \(\wedge\) symbol. For example, \(y_x \wedge z_x \wedge x'\) denotes the event that \(Y_x = y\), \(Z_x = z\) and \(X = x'\). The probability \(P(y_x \wedge z_x \wedge x')\) is the probability of the counterfactual event. Note that primes do not differentiate variables, instead they are used to differentiate between values i.e., \(x\) is a different value from \(x'\) and they are both different from \(x''\) but all three are value assignments of the random variable \(X\). If the subscript of each variable in the conjunction is the same, the counterfactual probability simply reduces to an interventional distribution.
  
\begin{figure}[ht]
  \centering
  \begin{subfigure}[t]{0.25\textwidth}
    \begin{center}
      \begin{tikzpicture}[scale=1.8]
        \node[obs = {X}] at (0,0) {\vphantom{0}};
        \node[obs = {W}] at (0,-1) {\vphantom{0}};
        \node[obs = {D}] at (1,0) {\vphantom{0}};
        \node[obs = {Z}] at (1,-1) {\vphantom{0}};
        \node[obs = {Y}] at (0.5,-2) {\vphantom{0}};
        \node[inner sep = 5pt, circle] at (0,-2.75) {\vphantom{0}};
        \draw [->] (X) -- (W);
        \draw [->] (W) -- (Y);
        \draw [->] (D) -- (Z);
        \draw [->] (Z) -- (Y);
        \draw[<->,dashed] (X) to[bend right=45] (Y);
      \end{tikzpicture}
    \end{center}
    \caption{A causal diagram.}
    \label{fig:parrallelexampleoriginal}
  \end{subfigure}
  \hfill
  \begin{subfigure}[t]{0.70\textwidth}
    \begin{center}
      \begin{tikzpicture}[scale=1.8]
        \node[obs = {X}] at (0,0) {\vphantom{0}};
        \node[obs = {W}] at (0,-1) {\vphantom{0}};
        \node[obs = {D}] at (1,0) {\vphantom{0}};
        \node[obs = {Z}] at (1,-1) {\vphantom{0}};
        \node[obs = {Y}] at (0.5,-2) {\vphantom{0}};
        \draw [->] (X) -- (W);
        \draw [->] (W) -- (Y);
        \draw [->] (D) -- (Z);
        \draw [->] (Z) -- (Y);
        \node[fixed = x, color=blue] at (2,0) {\vphantom{0}};
        \node[obs = {W_x}, color=blue] at (2,-1) {\vphantom{0}};
        \node[obs = {D_x}, color=blue] at (3,0) {\vphantom{0}};
        \node[obs = {Z_x}, color=blue] at (3,-1) {\vphantom{0}};
        \node[obs = {Y_x}, color=blue] at (2.5,-2) {\vphantom{0}};
        \draw [->] (x) -- (W_x);
        \draw [->] (W_x) -- (Y_x);
        \draw [->] (D_x) -- (Z_x);
        \draw [->] (Z_x) -- (Y_x);
        \node[obs = {X_d}, color=red] at (4,0) {\vphantom{0}};
        \node[obs = {W_d}, color=red] at (4,-1) {\vphantom{0}};
        \node[fixed = {d}, color=red] at (5,0) {\vphantom{0}};
        \node[obs = {Z_d}, color=red] at (5,-1) {\vphantom{0}};
        \node[obs = {Y_d}, color=red] at (4.5,-2) {\vphantom{0}};
        \draw [->] (X_d) -- (W_d);
        \draw [->] (W_d) -- (Y_d);
        \draw [->] (d) -- (Z_d);
        \draw [->] (Z_d) -- (Y_d);
        \node[lat = {U1}{U_D}] at (2,0.625) {\vphantom{0}};
        \node[lat = {U2}{U}] at (2,1.25) {\vphantom{0}};
        \node[lat = {U3}{U}] at (2.5,-2.75) {\vphantom{0}};
        \node[lat = {U4}{U_Z}] at (3.5,-1.625) {\vphantom{0}};
        \node[lat = {U5}{U_W}] at (1.5,-0.475) {\vphantom{0}};
        \draw [->,dashed] (U1) to[bend right=15] (D);
        \draw [->,dashed] (U1) to[bend left=15] (D_x);
        \draw [->,dashed] (U2) to[bend right=22.5] (X);
        \draw [->,dashed] (U2) to[bend left=22.5] (X_d);
        \draw [->,dashed] (U3) to[bend left=15] (Y);
        \draw [->,dashed] (U3) -- (Y_x);
        \draw [->,dashed] (U3) to[bend right=15] (Y_d);
        \draw [->,dashed] (U4) to[bend left=15] (Z);
        \draw [->,dashed] (U4) -- (Z_x);
        \draw [->,dashed] (U4) to[bend right=15] (Z_d);
        \draw [->,dashed] (U5) to[bend right=15] (W);
        \draw [->,dashed] (U5) -- (W_x);
        \draw [->,dashed] (U5) to[bend left=10] (W_d);
      \end{tikzpicture}
    \end{center}
    \caption{A parallel worlds graph of (a) for \(y_x \wedge x' \wedge z_d \wedge d\). Colors are used here to distinguish the observed and fixed nodes that belong to different parallel worlds: black for the non-interventional world, blue for the world induced by \(\doo(X = x)\), and red for the world induced by \(\doo(D = d)\). Note that node \(U\) is drawn twice for clarity due to its many endpoints.}
    \label{fig:parrallelexampleworlds}
  \end{subfigure}
  \caption{An example causal diagram and a corresponding parallel worlds graph.}
  \label{fig:parallelexample}
  \end{figure}
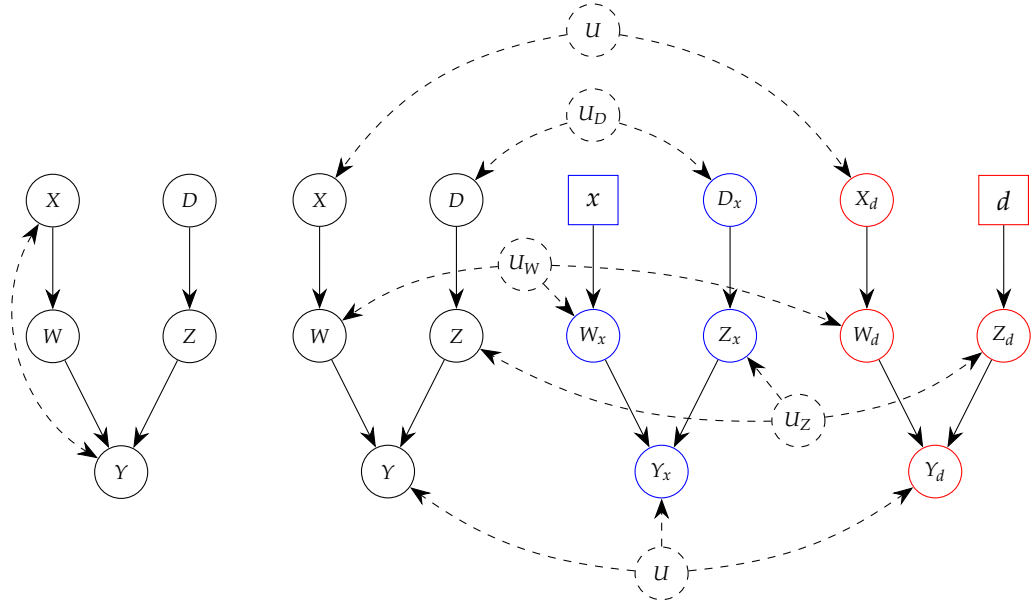

  Each counterfactual conjunction is associated with multiple \dfn{parallel worlds}, each induced by a unique combination of subscripts that appears in the conjunction. A \dfn{parallel worlds graph} of the conjunction is obtained by combining the graphs of the submodels induced by interventions such that the latent variables are shared. The simplest version of a parallel worlds graph is a twin network graph, contrasting two alternative worlds \citep{balke1994a, balke1994b, avin2005pathspecific}. As a more complicated example, consider the counterfactual conjunction \(\gamma = y_x \wedge x' \wedge z_d \wedge d\). In simpler terms, this conjunction states that \(Y\) takes the value \(y\) under the intervention \(\doo(X = x)\), \(Z\) takes the value \(z\) under the intervention \(\doo(D = d)\), and \(X\) and \(D\) take the values \(x'\) and \(d\), respectively, when no intervention took place. Importantly, this conjunction induces three distinct parallel worlds: the non-interventional (or observed) world, a world where \(X\) was intervened on, and a world where \(D\) was intervened on. For instance, if the graph in Figure~\ref{fig:parrallelexampleoriginal} depicts the original causal model over the variables \(Y, X, Z, W\) and \(D\), then Figure~\ref{fig:parrallelexampleworlds} shows the corresponding parallel worlds graph for \(\gamma\), where each distinct world is represented by its own set of copies of the original variables. In Figure~\ref{fig:parrallelexampleworlds}, \(U\) corresponds to the bidirected edge between \(X\) and \(Y\) in Figure~\ref{fig:parrallelexampleoriginal}, and the other \(U\)-variables are the individual error terms of each observed variable, that are not drawn when they have only one child in Figure~\ref{fig:parrallelexampleoriginal}.
  
  Note that instead of random variables, some nodes in the parallel worlds graph now depict fixed values as assigned by the interventions in the conjunction. This is a crucial aspect when d-separation statements are considered between counterfactual variables in the parallel worlds graph, as a backdoor path through a fixed value is not open. Furthermore, not every variable is necessarily unique in a parallel worlds graph, making it possible to obtain misleading results if d-separation is used to infer conditional independence relations between counterfactual variables. For instance, if we consider the counterfactual variables \(Y_x\), \(D_x\) and \(Z\) in a causal model whose diagram is the graph shown in Figure~\ref{fig:parrallelexampleoriginal}, then \(Y_x\) is independent of \(D_x\) given \(Z\), even though \(Y_x\) is not d-separated from \(D_x\) in the corresponding parallel worlds graph of Figure~\ref{fig:parrallelexampleworlds}. This conditional independence holds because \(Z\) and \(Z_x\) are in fact the same counterfactual variable. To overcome this problem, the parallel worlds graph must be further refined into the \dfn{counterfactual graph} where every variable is unique, which we will discuss in the following sections in more detail. For causal diagrams and counterfactual graphs, \(V(G)\) denotes the set of observable random variables not fixed by interventions and \(v(G)\) denotes the corresponding set of value assignments.
  
  The following operations are defined for counterfactual conjunctions and sets of counterfactual variables: \(\mathrm{sub}(\cdot)\) returns the set of subscripts, \(\mathrm{var}(\cdot)\) returns the set of (non-counterfactual) variables, and \(\mathrm{ev}(\cdot)\) returns the set of values (either fixed by intervention or observed). For example, consider again the conjunction \(\gamma = y_x \wedge x' \wedge z_d \wedge d\). Now, \(\mathrm{sub}(\gamma) = \{x,d\}\), \(\mathrm{var}(\gamma) = \{Y, X, Z, D\}\) and \(\mathrm{ev}(\gamma) = \{y, x, x',z,d\}\). Finally, \(\mathrm{val}(\cdot)\) is the value assigned to a given counterfactual variable, e.g., \(\mathrm{val}(y_x) = y\). The notation \(y_{\+ x..}\) denotes a counterfactual variable derived from \(Y\) with the value assignment \(y\) in a submodel \(M_{\+ x \cup \+ z}\) where \(\+ Z \subseteq \+ V \setminus \+ X\) is arbitrary.
  
  The symbol \(P_*\) is used to denote the set of all interventional distributions of a causal model \(M\) over a set of observed variables \(\+ V\), i.e., 
  \[
  P_* = \{P_{\+x} \mid \+ x \text{ is any value assignment of }\+ X \subseteq \+ V\}
  \]
  In the following sections, we consider identifiability of counterfactual queries in terms of \(P_*\). In essence, this means that a counterfactual probability distribution \(P(\gamma)\) is identifiable if it can be expressed using purely interventional and observational probabilities of the given causal model.
  
  \subsection{Example on identifiability of a counterfactual query} \label{sec:example}
  
  We consider the identifiability of the conditional counterfactual query \(P(y_x|z_x \wedge x')\) from \(P_*\) in the graph depicted in Figure~\ref{fig:graphG}. This graph could for instance depict the effect of an applicant's education (\(X\)) on work experience (\(Z\)) and a potential hiring decision (\(Y\)) by a company. Our counterfactual query could then consider the statement ``what is the probability to be hired if the applicant's education level was changed to \(x\), given that their work experience under the same intervention was \(z\) and when in reality their education level was \(x'\)''. In this example, we will not rely on any identifiability algorithms. Instead, we can derive a formula for the counterfactual query as follows:
  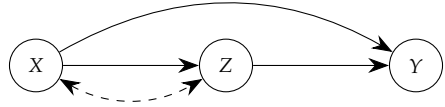
\begin{figure}[ht]
  \begin{center}
    \begin{tikzpicture}[scale=2.5]
      \node[obs = {X}] at (0,0) {\vphantom{0}};
      \node[obs = {Z}] at (1,0) {\vphantom{0}};
      \node[obs = {Y}] at (2,0) {\vphantom{0}};
      \draw [->] (X) -- (Z);
      \draw [->] (X) to [bend left=30] (Y);
      \draw [->] (Z) edge (Y);
      \draw[<->,dashed] (X) to[bend right=30] (Z);
    \end{tikzpicture}
  \end{center}
  \caption{A graph for the example on identifiability of a conditional counterfactual query \(P(y_x|z_x \wedge x')\).}
  \label{fig:graphG}
  \end{figure}
  
  \begin{align*}
    P(y_x|z_x \wedge x') 
      &= \frac{P(y_x \wedge z_x \wedge x')}{\sum_{y}P(y_x \wedge z_x \wedge x')} & \\
      &= \frac{P(y_{xz} \wedge z_x \wedge x')}{\sum_{y}P(y_{xz} \wedge z_x \wedge x')} & (\text{composition})\\
      &= \frac{P(y_{xz} | z_x \wedge x')P(z_x \wedge x')}{\sum_{y}P(y_{xz} | z_x \wedge x')P(z_x \wedge x')} & \\
      &= \frac{P(y_{xz})P(z_x \wedge x')}{P(z_x \wedge x')\sum_y P(y_{xz})} & (\text{independence restrictions}) \\
      &= P(y_{xz}) \\
      &= P_{xz}(y)
  \end{align*}
  Thus, we find that the answer to our initial question is simply the probability of hiring if the applicant's education level and work experience were changed to \(x\) and \(z\), respectively. In the above derivation, we used the notions of composition and independence restrictions \citep{holland1986, pearl1995, halpern1998, pearl2009}. Composition is one of the axioms of counterfactuals stating that if a variable is forced to a value that it would have taken without the intervention, then the intervention will not affect other variables in the system. In this case, intervention setting \(Z_x\) to \(z\) has no effect on \(Y_x\) because we have observed \(Z_x = z\), thus we can add \(Z\) to the intervention set of \(Y_x\). Independence restrictions state if the observed parents of a variable are intervened on, then the counterfactual is independent of any other observed variable when their parents are also held fixed, if there are no paths between the variables via latent variables. In this case \(Y_{x,z}\) is independent of \(Z_x\) and \(X\) because there is no path via latent variables connecting \(Y\) to \(Z\) or \(X\) in \(G\). 
  
  In this example, the interventional distribution \(P_{x,z}(y)\) can be further identified from the observed joint distribution \(P(x,z,y)\) as \(P(y|x,z)\) via the second rule of do-calculus by noting that \(Y\) is d-separated from \(X\) and \(Z\) in the graph when the outgoing edges of \(X\) and \(Z\) are removed. Thus, the answer to our initial question can be further refined into the probability of hiring among applicants with education level \(x\) and work experience \(z\). The \pkg{cfid} package provides this kind of identification pipeline from the counterfactual level down to the lowest possible level in the causal hierarchy.
  
  \section{Algorithms for identifying counterfactual queries} \label{sec:algo}
  
  Manual identification of counterfactuals is challenging and more nuanced than identification of interventional distributions due to fixed values and non-uniqueness of counterfactual variables in the parallel worlds graph. Therefore, identification of a counterfactual query can be achieved in several ways. First, we may find that the query is identifiable and thus we can express it in terms of purely interventional distributions. In contrast, we may find that the query is not identifiable, meaning that is not possible to represent it in terms of purely interventional distributions. Alternatively, we may find that the query is \dfn{inconsistent} meaning that the same counterfactual variable has been assigned at least two different values in the conjunction, and thus the query is identified as a zero-probability event. For example, suppose we are tasked with identifying \(P(y_x, y'_z)\) but we find that \(Y_x\) and \(Y_z\) are actually the same variable, and thus cannot attain two different values \(y\) and \(y'\) simultaneously. For conditional counterfactual queries, there is also a fourth option where the query is undefined if the conditioning conjunction is inconsistent.
  
  Algorithmic identification of interventional distributions takes advantage of the so-called \dfn{C-component factorization} \citep{tian2002general, shpitser2006id} which also plays a key role in the identification of counterfactual queries. The \dfn{maximal C-components} of a causal diagram are obtained by partitioning the vertices \(\+ V\) related to observed variables of the graph such that two vertices \(A, B \in B\) in the same partition are connected by a path with edges into \(A\) and \(B\) where every node on the path in \(\+ V\) except \(A\) and \(B\) is a collider, and \(A\) and \(B\) are not connected to any other partitions via such paths. Maximal C-components are defined analogously for parallel worlds graphs and counterfactual graphs with the restriction that we do not consider vertices that correspond to fixed values to belong to any C-component. The set of maximal C-components of a DAG \(G\) is denoted by \(C(G)\). As an example, the maximal C-components of the graph of Figure~\ref{fig:parrallelexampleworlds} are \(\{X, X_d, Y, Y_x, Y_d\}\), \(\{D, D_x\}\), \(\{Z, Z_x, Z_d\}\), and \(\{W, W_x, W_d\}\).
  
  We recall the ID* and IDC* algorithms of \citet{shpitser2007} which are depicted in Figures~\ref{fig:idstaralgo} and \ref{fig:idcstaralgo} for identifying counterfactual queries and conditional counterfactual queries, respectively. Both algorithms are sound and complete \citep[Theorems 26 and 31]{shpitser2008}, meaning that when they succeed in identifying the query, the expression returned is equal to \(P(\gamma)\) or \(P(\gamma|\delta)\), respectively, and when they fail, the query is not identifiable. We aim to characterize the operation of these algorithms on an intuitive level and provide line-by-line examples of their operation via examples.
  
  \begin{figure}[ht]
  \begin{center}
    \begin{tabular}{l}
      function \textbf{ID*}(\(G\), \(\gamma\)) \\
      INPUT: \(G\) a causal diagram, \(\gamma\)  a conjunction of counterfactual events\\
      OUTPUT: an expression for \(P(\gamma)\) in terms of \(P_*\), or \textbf{FAIL}
    \end{tabular}
    \begin{minipage}{.80\textwidth}
    \vspace*{0.25cm}
    \begin{enumerate}
      \item if \(\gamma = \emptyset\), return 1
      \item if \((\exists x_{x'..} \in \gamma)\), return 0
      \item if \((\exists x_{x..} \in \gamma)\), return \textbf{ID*}\((G, \gamma \setminus \{x_{x..}\})\)
      \item \((G', \gamma')\) = \textbf{make-cg}\((G, \gamma)\)
      \item if \(\gamma'\) = \textbf{INCONSISTENT}, return 0
      \item if \(C(G') = \{\+ S^1, \ldots, \+ S^k\}\), return \(\sum_{V(G')\setminus \gamma'} \prod_{i=1}^k\)\textbf{ID*}\((G, \+ s^i_{v(G')\setminus \+ s^i})\)
      \item if \(C(G') = \{\+ S\}\), then
      \addtocounter{enumi}{1}
      \begin{enumerate}[label=\arabic*.,start=\value{enumi}]
        \item if \((\exists \+x, \+x')\) s.t. \(\+ x \neq \+ x', \+ x \in \mathrm{sub}(\+ S), \+ x' \in \mathrm{ev}(\+ S)\), throw \textbf{FAIL}
        \item else, let \(\+ x =  \cup\,\mathrm{sub}(\+ S)\), return \(P_{\+ x}(\mathrm{var}(\+ S))\).
      \end{enumerate}
    \end{enumerate}
    \end{minipage}
  \end{center}
  \caption{Counterfactual identification algorithm ID* by \citet{shpitser2007}.}
  \label{fig:idstaralgo}
  \end{figure}
  
  We begin by describing the ID* algorithm. On line~1, we check for an empty conjunction, which by convention has probability 1. Line~2 checks for direct inconsistencies meaning counterfactual variables that are intervened on but simultaneously observed to have a different value than the intervention. Such counterfactuals violate the Axiom of Effectiveness \citep{pearl2009}, and if found, we return probability 0. Line~3 removes tautological counterfactuals from the conjunction meaning counterfactuals where the variable was observed to have the value it was forced to take via intervention. Line~4 calls the make-cg algorithm to construct the counterfactual graph \(G'\) and the corresponding conjunction \(\gamma'\) where some counterfactual variables may have been relabeled due to equivalence between counterfactual variables. We leave the details of the make-cg algorithm and the related core results to the supplementary material. In summary, the output \(G'\) of make-cg is a refined version of the parallel worlds graph of \(G\) and \(\gamma\), where each counterfactual variable is unique. Similarly, if some variables in \(\gamma\) were found to be equivalent, then those variables are replaced in \(\gamma'\) by their new representatives in \(G'\). If as a result of this operation the refined conjunction \(\gamma'\) is now inconsistent, we again return probability 0. The next two lines take advantage of the C-component factorization of the counterfactual graph \(G'\), analogously to the ID algorithm. If there is more than one maximal C-component of \(G'\), then we proceed to line~6 where the original query is decomposed into a set of subproblems, each of which we again call ID* for. Note that the sets \(\+ S^i\) are sets of counterfactual variables, but we may interpret them as counterfactual conjunctions in the subsequent recursive calls. Similarly, we may interpret \(\gamma'\) as a set of counterfactual variables when carrying out the outermost summation over the possible values of the counterfactual variables in \(V(G')\setminus \gamma'\). In cases where a set \(\+ S^i\) contains counterfactual variables, the intervention \(\doo(v(G') \setminus \+ s^i)\) should be understood as merging of the subscripts, e.g., if \(\+ S^i = \{Y_x\}\) and \(V(G') \setminus \+ S^i = \{Z\}\), and \(Y_x\) has the value \(y\) in \(\gamma'\), then \(\+ s^i_{v(G')\setminus \+ s^i} = y_{x,z}\).
  
  If there is only one C-component, we enter line~7 that serves as the base case. There are now only two options. If there is an inconsistent value assignment on line~8 such that at least one of the values is in the subscript, then the query is not identifiable, and we fail. If there is no such conflict, we can take the union of all the subscripts in \(\gamma'\) and return their effect on the variables in \(\gamma'\) on line~9.
  
  \begin{figure}[ht]
  \begin{center}
    \begin{tabular}{l}
      function \textbf{IDC*}(\(G\), \(\gamma\), \(\delta\)) \\
      INPUT: \(G\) a causal diagram, \(\gamma, \delta\) conjunctions of counterfactual events\\
      OUTPUT: an expression for \(P(\gamma|\delta)\) in terms of \(P_*\), or \textbf{FAIL}, or \textbf{UNDEFINED}
    \end{tabular}
    \begin{minipage}{.80\textwidth}
    \vspace*{0.25cm}
    \begin{enumerate}
      \item if \textbf{ID*}\((G, \delta) = 0\), return \textbf{UNDEFINED}
      \item \((G', \gamma' \wedge \delta')\) = \textbf{make-cg}\((G, \gamma \wedge \delta)\)
      \item if \(\gamma' \wedge \delta'\) = \textbf{INCONSISTENT}, return \(0\)
      \item if \((\exists y_{\+{x}} \in \delta')\) s.t. \((Y_\+{x} \independent \gamma') G'_{\underline{y_{\+{x}}}}\), return \textbf{IDC*}\((G,\gamma'_{y_{\+{x}}}, \delta' \setminus \{y_{\+{x}}\})\)
      \item else, let \(P'\) = \textbf{ID*}\((G, \gamma' \wedge \delta')\), return \(P'/P'(\delta)\)
    \end{enumerate}
    \end{minipage}
  \end{center}
  \caption{Conditional counterfactual identification algorithm IDC* by \citet{shpitser2007}.}
  \label{fig:idcstaralgo}
  \end{figure}
  
  In contrast, the IDC* algorithm is simpler, as it leverages the ID* algorithm. The consistency of the conditioning conjunction \(\delta\) is first confirmed on line~1, and if \(\delta\) is found to be inconsistent, then the conditional probability \(P(\gamma|\delta)\) is undefined, and we return. Line~2 applies the make-cg algorithm to the joint conjunction \(\gamma \wedge \delta\) to construct the corresponding counterfactual graph \(G'\) and the restructured version of the conjunction, \(\gamma' \wedge \delta'\). If \(\gamma' \wedge \delta'\) was found to be inconsistent, we return probability 0 on line~3. Line~4 takes advantage of conditional independence relations implied by the counterfactual graph \(G'\) and the second rule of do-calculus to add variables as interventions to \(\gamma'\) by removing them from \(\delta'\). If the necessary d-separation holds, we initiate a recursive call to IDC* again. Finally on line~5, if no more variables can be removed from \(\delta'\), we simply apply the ID* algorithm to the joint conjunction \(\gamma' \wedge \delta'\) and obtain the identifying functional as a standard conditional probability from the distribution returned by ID*.
  
  \subsection{Examples on the identifiability algorithm} \label{sec:id_examples}
  
  We recall the counterfactual conjunction \(\gamma = y_x \wedge x' \wedge z_d \wedge d\) from Section~\ref{sec:notation} and describe how the ID* algorithm operates when applied to \(P(\gamma)\) in the graph of Figure~\ref{fig:parrallelexampleoriginal}, which we will label as \(G\) in the context of this example. We start from line~1 and continue to line~2 as \(\gamma\) is not an empty conjunction. On line~2, we note that \(\gamma\) does not contain any inconsistencies, similarly on line~3 we see that \(\gamma\) does not contain any tautological statements. Thus, we reach line~4 and apply the make-cg algorithm to obtain the counterfactual graph \(G'\) and the modified conjunction \(\gamma'\).
  
  We describe the operation of the make-cg algorithm in this instance. The goal is to determine which variables in the parallel worlds graph of Figure~\ref{fig:parrallelexampleworlds} represent the same variable. We consider all variable pairs in a topological order of \(G\) that originate from the same non-counterfactual variable in \(G\). First, we can conclude that \(X\) and \(X_d\) are the same variable, as they have the same functional mechanisms and the same parent \(U\). By the same argument, \(D\) and \(D_x\) are the same variable with the common parent \(U_D\). The fixed variables \(x\) and \(d\) cannot be merged with the other \(X\)-derived variables and \(D\)-derived variables, respectively, as their functional mechanisms are different. Next, we merge \(W\) and \(W_d\) because their \(X\)-derived parents (\(X\) and \(X_d\)) were found to be the same and they have the same parent \(U_W\). However, \(W_X\) cannot be merged with the other two \(W\)-derived variables, because \(X\) (and thus \(X_d\)) was observed to attain the value \(x'\) in \(\gamma\), but \(x\) has the value \(x\) as fixed by the intervention. In contrast, we can merge the triplet \(Z\), \(Z_x\) and \(Z_d\), because their \(D\)-derived parents attain the same value, and they have the same parent \(U_Z\). The intuition is that because the \(U\)-variables are shared between worlds, intervention and observation have the same effect if the observed values agree with the values fixed by intervention. This is a consequence of the Axiom of Composition as was considered in the example of Section~\ref{sec:example}. Finally, we consider the \(Y\)-derived variables and merge \(Y_x\) and \(Y_d\) because their \(Z\)-derived parents are the same, their \(W\)-derived parents are the same, and they have the same parent \(U\). The variable \(Y_x\) cannot be merged with the other two, because its \(W\)-derived parent \(W_x\) was not the same variable as \(W\) and \(W_d\). 
  
  Consequently, we must choose a name for each merged variable. This choice is arbitrary and plays no role in the correctness of the algorithm; the difference is purely notational. In this example, we pick the original name with the fewest subscripts to represent the merged variable, i.e., \(X\) represents the merged pair \(X, X_d\), \(Z\) represents the merged triplet \(Z, Z_x, Z_d\), \(W\) represents the merged pair \(W, W_d\) and finally \(Y\) represents the merged pair \(Y, Y_d\). Note that because the \(Z\)-derived variables were all merged but \(d\) was not merged with \(D\) and \(D_x\), we essentially have two \(D\)-derived parents for the merged \(Z\). In such scenarios, we simply omit the fixed version of the parent variable from the graph, because this scenario may only arise if the parent variables were found to have the same value, thus their role in the functional mechanisms of their children is identical. Lastly, we may restrict our attention to those counterfactual variables that are ancestral to the query \(\gamma\) in this merged graph, which are \(x, W_x, Y_x, Z, D, X\) and \(U\)
  
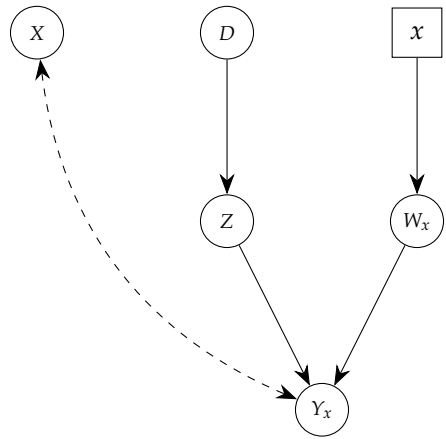
\begin{figure}[ht]
    \begin{center}
      \begin{tikzpicture}[scale=2.5]
        \node[obs = {X}] at (0,0) {\vphantom{0}};
        \node[obs = {D}] at (1,0) {\vphantom{0}};
        \node[obs = {Z}] at (1,-1) {\vphantom{0}};
        \draw [->] (D) -- (Z);
        \node[fixed = x] at (2,0) {\vphantom{0}};
        \node[obs = {W_x}] at (2,-1) {\vphantom{0}};
        \node[obs = {Y_x}] at (1.5,-2) {\vphantom{0}};
        \draw [->] (x) -- (W_x);
        \draw [->] (W_x) -- (Y_x);
        \draw [->] (Z) -- (Y_x);
        \draw [<->,dashed] (X) to[bend right=30] (Y_x);
      \end{tikzpicture}
    \end{center}
    \caption{Counterfactual graph \(G'\) for \(y_x \wedge x' \wedge z_d \wedge d\) of the graph of Figure~\ref{fig:parrallelexampleoriginal}.}
    \label{fig:cgid}
  \end{figure}

  Thus, we obtain the counterfactual graph \(G'\) for \(\gamma\) depicted in Figure~\ref{fig:cgid} using once again the convention that unobserved variables with only one child are not drawn. As a result of the variable merges, we also update our original conjunction \(\gamma\) with references to the merged variables to obtain \(\gamma' = y_x \wedge x' \wedge z \wedge d\). The new conjunction \(\gamma'\) is not inconsistent on line~5, and thus we continue.
  
  On line~6 we first determine the maximal C-components of the counterfactual graph \(G'\) which are \(\{X, Y_x\}\), \(\{Z\}\), \(\{W_x\}\) and \(\{D\}\). By the C-component factorization we have that 
  \begin{equation} \label{eq:idfactors}
    P(y_x \wedge x' \wedge z \wedge d) = \sum_{w} P(y_{x,z,w,d} \wedge x'_{z,w,d})P(z_{y,x,w,d})P(w_{x,y,z,d})P(d_{y,x,z,w}),
  \end{equation}
  which means that we launch four recursive calls to ID* to identify each of the terms in the right-hand side expression. We will consider the last three terms first as they result in a similar simple path through the algorithm. For each of these terms, the counterfactual graph will contain a single non-fixed vertex (\(Z_{y,x,w,d}\), \(W_{x,y,z,d}\) and \(D_{y,x,z,w}\), respectively). Because the conjunctions are not empty, there are no inconsistencies or tautologies, and only a single C-component, we end up on line~7 in each case. None of the terms contain value assignments that would conflict with the subscript and thus each term is identified as an interventional distribution on line~9. Note that when line~7 is reached, redundant subscripts should be removed, i.e., those subscript variables that are not ancestors of the counterfactual variables in \(\gamma'\) in the counterfactual graph \(G'\). Otherwise, a conflict may be found erroneously on line~8. This operation was not formally included in the algorithm by \citet{shpitser2007}, but nonetheless carried out in a running example by \citet{shpitser2008}. Thus, \(P(z_{y,x,w,d}) = P_{d}(z)\), \(P(w_{x,y,z,d}) = P_{x}(w)\) and \(P(d_{y,x,z,w}) = P(d)\). For the first term \(P(y_{x,z,w,d} \wedge x'_{z,w,d})\), the only difference is that the counterfactual graph has two non-fixed vertices, but the outcome is the same and we end up on line~7 due to the single C-component containing \(Y_{x,z,w,d}\) and \(X_{z,w,d}\). There are no conflicts this time either, and we obtain \(P(y_{x,z,w,d} \wedge x'_{z,w,d}) = P_{w,z}(y,x')\). Thus, we obtain the identifying functional of the counterfactual query:
  \[
    P(y_x \wedge x' \wedge z_d \wedge d) = \sum_{w} P_{w,z}(y,x')P_{d}(z)P_{x}(w)P(d).
  \]
  
  Next, we will consider an example that causes a conflict at line~7 resulting in a non-identifiable counterfactual query. Suppose that we also have an edge from \(X\) to \(Y\) in the graph of Figure~\ref{fig:parrallelexampleoriginal} and we wish to identify the same counterfactual query \(P(y_x \wedge x' \wedge z \wedge d)\) as in the previous example in this modified graph. The ID* algorithm proceeds similarly as in the previous example up to line~4 where we obtain a slightly different counterfactual graph, which is the graph of Figure~\ref{fig:cgid}, but with the corresponding extra edge from \(X\) to \(Y_x\). Thus, the algorithm proceeds similarly to line~6, where the C-component factorization is the same as \eqref{eq:idfactors}. The last three terms are still identifiable, but this time the first term \(P(y_{x,z,w,d} \wedge x'_{z,w,d})\) is problematic. On line~7 after removing redundant interventions, the term takes the form \(P(y_{x,z,w} \wedge x'_{z,w})\) which now contains a conflict, because \(x\) appears in the subscript but \(x'\) is observed at the same time, resulting in non-identification on line~8.

\begin{figure}[ht]
  \centering
  \begin{subfigure}{0.48\textwidth}
    \begin{center}
      \begin{tikzpicture}[scale=2.5]
        \node[obs = {X}] at (0,0) {\vphantom{0}};
        \node[fixed = {x}] at (0,-1) {\vphantom{0}};
        \node[obs = {Z}] at (1,0) {\vphantom{0}};
        \node[obs = {Z_x}] at (1,-1) {\vphantom{0}};
        \node[obs = {Y}] at (2,0) {\vphantom{0}};
        \node[obs = {Y_x}] at (2,-1) {\vphantom{0}};
        \node[lat = {U1}{U}] at (0.5,-0.5) {\vphantom{0}};
        \node[lat = {U2}{U_Y}] at (1.5,-0.5) {\vphantom{0}};
        \draw [->] (X) -- (Z);
        \draw [->] (x) -- (Z_x);
        \draw [->] (X) to [bend left=30] (Y);
        \draw [->] (x) to [bend right=30] (Y_x);
        \draw [->] (Z) -- (Y);
        \draw [->] (Z_x) -- (Y_x);
        \draw [->,dashed] (U1) -- (X);
        \draw [->,dashed] (U1) -- (Z);
        \draw [->,dashed] (U1) -- (Z_x);
        \draw [->,dashed] (U2) -- (Y);
        \draw [->,dashed] (U2) -- (Y_x);
      \end{tikzpicture}
    \end{center}
    \caption{Parallel worlds graph for \(y_x \wedge z_x \wedge x'\) (the counterfactual graph).}
    \label{fig:parallelidc1}
  \end{subfigure}
  \hfill
  \begin{subfigure}{0.48\textwidth}
    \begin{center}
      \begin{tikzpicture}[scale=2.5]
        \node[obs = {X}] at (0,0) {\vphantom{0}};
        \node[fixed = {x}] at (0,-1) {\vphantom{0}};
        \node[obs = {Z}] at (1,0) {\vphantom{0}};
        \node[fixed = {z}] at (1,-1) {\vphantom{0}};
        \node[obs = {Y}] at (2,0) {\vphantom{0}};
        \node[obs = {Y_{x,z}}] at (2,-1) {\vphantom{0}};
        \node[lat = {U1}{U}] at (0.5,-0.5) {\vphantom{0}};
        \node[lat = {U2}{U_Y}] at (1.5,-0.5) {\vphantom{0}};
        \draw [->] (X) -- (Z);
        \draw [->] (X) to [bend left=30] (Y);
        \draw [->] (x) to [bend right=30] (Y_{x,z});
        \draw [->] (Z) -- (Y);
        \draw [->] (z) -- (Y_{x,z});
        \draw [->,dashed] (U1) -- (X);
        \draw [->,dashed] (U1) -- (Z);
        \draw [->,dashed] (U2) -- (Y);
        \draw [->,dashed] (U2) -- (Y_{x,z});
      \end{tikzpicture}
    \end{center}
    \caption{Parallel worlds graph for \(y_{x,z} \wedge x'\) (the counterfactual graph).}
    \label{fig:parallelidc2}
  \end{subfigure}
    \caption{Counterfactual graphs used during the derivation of \(P(y_x | z_x \wedge x')\).}
    \label{fig:parallelidc}
  \end{figure}
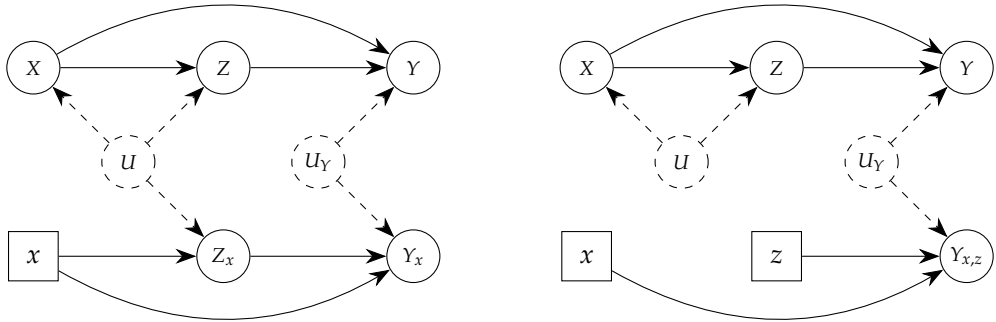

  
  We return to the example presented in Section~\ref{sec:example} and apply the IDC* algorithm to identify the counterfactual query \(P(y_x| z_x \wedge x')\) in the graph of Figure~\ref{fig:graphG}, which we will again refer to as \(G\) in the context of this example. We trace the application of IDC*\((G, y_x, z_x \wedge x')\). On line~1, the ID* algorithm is applied to \(z_x \wedge x'\), which is not identifiable, but also not inconsistent. Continuing to line~2, we apply the make-cg algorithm to construct the counterfactual graph \(G'\), which is shown in Figure~\ref{fig:parallelidc1}. First, the parallel worlds graph is constructed and \textbf{make-cg} proceeds to determine which variable pairs can be merged (see the Supplementary Material for details on the \textbf{make-cg} algorithm). 
  
  Because \(X\) was observed to have the value \(x'\), but the intervention for \(Z\) and \(Y\) has the value \(x\), we cannot merge \(X\) and \(x\). Similarly, the \(X\)-parent of \(Z\) in both worlds has a different value, meaning that \(Z\) and \(Z_x\) cannot be merged either. Finally, through the same reasoning, \(Y\) and \(Y_x\) will remain unmerged due to the difference in the \(Z\)-parent. Thus, the parallel worlds graph is the counterfactual graph \(G'\) in this instance. This also means that \(\gamma' = \gamma\) and \(\delta' = \delta\) in the output of make-cg.
  
  On line~3, we check for inconsistencies in \(y_x \wedge z_x \wedge x'\), but there are none. Next on line~4, we check whether either of the two variables in \(\delta'\) are d-separated from \(\gamma'\) when outgoing edges of that variable have been removed. We can see that \(X\) is not d-separated from \(Y_x\), because the path \(X \leftarrow U_1 \rightarrow Z_x \rightarrow Y_x\) is open in \(G'_{\underline X}\). However, \(Z_x\) is d-separated from \(Y_x\) in \(G'_{\underline{Z_x}}\) (note that \(x\) is fixed by intervention, and thus the path \(Z_x \leftarrow x \rightarrow Y_x\) is not an open backdoor path). Thus, line~4 adds an intervention on \(Z\) to \(Y_x\) because \(Y_x\) is a descendant of \(Z_x\) in \(G'\), and removes \(Z_x\) from \(\delta'\), and we call IDC*\((G', y_{x,z}, x')\).
  
  We now trace this new recursive call. Once again on line~1, ID* is not able to identify the effect, but is also not inconsistent. Next, we construct a new counterfactual graph \(G''\) for \(y_{x,z} \wedge x'\) as depicted in Figure~\ref{fig:parallelidc2}. Using similar reasoning as before, the make-cg algorithm is not able to merge any nodes this time either and thus the parallel worlds graph is the counterfactual graph. Again, this means that \(\gamma'' = \gamma'\) and \(\delta'' = \delta'\) in the output of make-cg. Line~3 checks again for inconsistencies in \(y_{x,z} \wedge x'\), but there are none. Thus, we arrive again on line~4, but this time \(X\) is d-separated from \(Y_{x,z}\) in \(G''_{\underline X}\). Now, \(Y_{x,z}\) is not a descendant of \(X\) in \(G''\) so no new intervention is added to \(Y_{x,z}\), and \(x'\) is removed from \(\delta''\). Because the conditioning \(\delta\)-argument of the next IDC* call is now empty, we can call ID* directly as ID*\((G, y_{x,z})\), but \(P(y_{x,z})\) is no longer a counterfactual quantity, but an interventional distribution and thus directly identifiable from \(P_*\) as \(P_{x,z}(y)\).
  
  We note the difference compared to the manual identification strategy we used in Section~\ref{sec:example} to obtain identifiability. Instead of using axioms of counterfactuals or independence restrictions explicitly, the ID* and IDC* algorithms take full advantage of the counterfactual graph and the conditional independence relations between the counterfactual variables implied by it.
  
  \section{Using the cfid package} \label{sec:package}
  
  The \pkg{cfid} package is available from CRAN at \url{https://cran.r-project.org/package=cfid} and can be obtained in R using the following commands:
  \begin{example}
  R> install.packages("cfid")
  R> library("cfid")
  \end{example}
  Development of \pkg{cfid} takes place on GitHub \url{https://github.com/santikka/cfid}.
  
  The main contributions of the \pkg{cfid} package are the implementations of the ID* and IDC* algorithms. The package also provides reimplementations of the ID and IDC algorithms for interventional distributions from the \pkg{causaleffect} package, but without relying on the \CRANpkg{igraph} \citep{igraph} package. In fact, \pkg{cfid} has no mandatory package dependencies or installation requirements. The \pkg{cfid} package provides its own text-based interface for defining graphs, which closely follows the syntax of the \pkg{dagitty} package, and also supports other external graph formats directly. Installation of the \pkg{igraph} and \pkg{dagitty} packages is optional and required only if the user wishes to import or export graphs using the aforementioned packages.
  
  The inclusion of the identifiability algorithms for interventional distributions enables a full identification pipeline. First, we determine the identifiability of a counterfactual query from the set of all interventional distributions, and then proceed to identify each interventional distribution that appears in the identifying functional of the counterfactual from the joint observed probability distribution of the causal model. The level of attempted identification can be specified by the user.
  
  \subsection{Defining causal diagrams} \label{sec:graphs}
  
  Causal diagrams (i.e., DAGs) in \pkg{cfid} are constructed via the function \code{dag}
  \begin{example}
  dag(x, u = character(0L)) 
  \end{example}
  where \code{x} is a single character string in a syntax analogous to the DOT language for GraphViz (and the \pkg{dagitty} package), and \code{u} is an optional character vector of variable names that should be considered unobserved in the graph. Internally, a semi-Markovian representation is always used for DAGs where each latent variable has at most two children, which is obtained from the input via the latent projection \citep{verma1990}.
  
  As an example, the graph of Figure~\ref{fig:graphG} can be constructed as follows:
  \begin{example}
  R> g <- dag("X -> Z -> Y; X -> Y; X <-> Z")
  \end{example}
  Above, individual statements are separated by a semicolon for additional clarity, but this is optional, and a space would suffice. More generally, the input of \code{dag} consists of statements of the form \(n_1 e_1 n_2 e_2 \cdots e_k n_k\) where each \(e_i\) symbol must be a supported edge type, i.e., \code{->}, \code{<-} or \code{<->}, and each \(n_i\) symbol must correspond to single node such as \code{X} or a subgraph such as \code{\{X, Y, Z\}} or \code{\{X ->\hphantom{,}Y\}}. Subgraphs are enclosed within curly braces, and they follow the same syntax as \code{x}. Subgraphs can also be nested arbitrarily. An edge of the form \code{X ->\hphantom{,}\{...\}} means that there is an edge from \code{X} to all vertices in the subgraph, and the interpretation for \code{<-} and \code{<->} is analogous. Individual statements in the graph definition can be separated by a semicolon, a space, or a new line. Commas can be used within subgraphs to distinguish vertices, but a space is sufficient.
  
  The same DAG can often be defined in many ways. For example, we could also define the graph of Figure~\ref{fig:graphG} using a subgraph construct as follows:
  \begin{example}
  R> g <- dag("X -> {Z, Y}; Z -> Y; X <-> Z")
  \end{example}
  We could also combine the outgoing edge of \code{Z} and the bidirected edge into a single statement:
  \begin{example}
  R> g <- dag("X -> {Z, Y}; X <-> Z -> Y;")
  \end{example}
  The edge from \code{Z} to \code{Y} could be defined in the subgraph as well:
  \begin{example}
  R> g <- dag("Z <-> X -> {Z -> Y}")
  \end{example}
  The output of \code{dag} is an object of class \code{"dag"} which is a square adjacency matrix of the graph, with additional attributes for the vertex labels and latent variables and a \code{print} method. Graph definitions that imply cycles or self-loops will raise an error. Examples of more complicated graph constructs can be found from the \pkg{cfid} package documentation for the \code{dag} function. Graphs using supported external formats can be converted to \code{"dag"} objects via the function \code{import\_graph}. Conversely, \code{"dag"} objects can be exported in supported external formats using the function \code{export\_graph}.
  
  \subsection{Defining counterfactual variables and conjunctions} \label{sec:cfs}
  
  Counterfactual variables are defined via the function \code{counterfactual\_variable} or its shorthand alias \code{cf}
  \begin{example}
  counterfactual_variable(var, obs = integer(0L), sub = integer(0L))
  cf(var, obs = integer(0L), sub = integer(0L))
  \end{example}
  The first argument \code{var} is a single character string naming the variable, e.g., \code{"Y"}. The second argument \code{obs} describes the value assignment as a single integer. The value of this argument does not describe the actual value taken by the variable, but simply the assignment level, meaning that \code{obs = 1} is a different value assignment than \code{obs = 0}, but the actual values that the counterfactual variable takes need not necessarily be 1 and 0. The idea is similar to the internal type of factors in R. Finally, \code{sub} defines the set of interventions as a named integer vector, where the actual values correspond to the intervention levels, and not actual values, analogous to \code{obs}. The output of \code{cf} is an object of class \code{"counterfactual\_variable"}.
  
  As an example, the counterfactual variables in \(\gamma = y_x \wedge x^\prime \wedge z_d \wedge d\) can be defined as follows:
  \begin{example}
  R> v1 <- cf(var = "Y", obs = 0L, sub = c(X = 0L))
  R> v2 <- cf(var = "X", obs = 1L)
  R> v3 <- cf(var = "Z", obs = 0L, sub = c(D = 0L))
  R> v4 <- cf(var = "D", obs = 0L)
  R> list(v1, v2, v3, v4)
  \end{example}
  \begin{example}
  [[1]]
  y_{x} 
  
  [[2]]
  x' 
  
  [[3]]
  z_{d} 
  
  [[4]]
  d 
  \end{example}
  The \code{print} method for \code{"counterfactual\_variable"} objects mimics the notation used in this paper in LaTeX syntax.
  
  Individual \code{"counterfactual\_variable"} objects can be combined into a counterfactual conjunction via the function \code{counterfactual\_conjunction} or its shorthand alias \code{conj}. This function takes arbitrarily many \code{"counterfactual\_variable"} objects as input. The output of \code{conj} is an object of class \code{"counterfactual\_conjunction"}.
  \begin{example}
  R> c1 <- conj(v1, v2, v3, v4)
  R> c1
  \end{example}
  \begin{example}
  y_{x} /\ x' /\ z_{d} /\ d 
  \end{example}
  Alternatively, the \code{`+`} operator can be used to build conjunctions from counterfactual variables or conjunctions.
  \begin{example}
  R> c2 <- v1 + v2
  R> c3 <- v3 + v4
  R> c2
  R> c3
  R> c2 + c3
  \end{example}
  \begin{example}
  y_{x} /\ x'
  z_{d} /\ d 
  y_{x} /\ x' /\ z_{d} /\ d 
  \end{example}
  The subset operator \code{`[`} is supported for counterfactual conjunctions
  \begin{example}
  R> c1[c(1, 3)]
  \end{example}
  \begin{example}
  y_{x} /\ z_{d}
  \end{example}
  Just as the \code{cf} function, the \code{print} method for \code{"counterfactual\_conjunction"} objects mimics the formal notation of using the \(\wedge\) symbol to separate individual statements, but this symbol can also be changed by the user.
  
  \subsection{Identifying counterfactual queries} \label{sec:id}
  
  Identification of counterfactual queries is carried out by the function \code{identifiable}
  \begin{example}
  identifiable(g, gamma, delta = NULL, data = "interventions") 
  \end{example}
  where \code{g} is a causal diagram defined by the function \code{dag}, \code{gamma} is the counterfactual conjunction \(\gamma\) as a \code{"counterfactual\_conjunction"} object describing the counterfactual query \(P(\gamma)\) to be identified, \code{delta} is an optional argument also of class \code{"counterfactual\_conjunction"} that should be provided if identification of a conditional counterfactual \(P(\gamma|\delta)\) is desired instead. Finally, \code{data} defines the available probability distributions for identification. The default value \code{"interventions"} means that identification is carried out to the intervention level, i.e., by using only the set of all interventional distributions \(P_*\). The alternatives are \code{"observations"}, where only the joint observed probability distribution \(P(\+ v)\) is available, and \code{"both"} where both \(P_*\) and \(P(\+ v)\) are available, and identification in terms of \(P(\+ v)\) is prioritized.
  
  We reassess the identifiability examples of Section~\ref{sec:id_examples} using the \pkg{cfid} package. The conjunction of the query \(\gamma\) for the first two examples is the same as \code{c1} in the previous section. We define the graphs for the identifiable case in Figure~\ref{fig:parrallelexampleoriginal} and the non-identifiable case with the additional edge from \(X\) to \(Y\):
  \begin{example}
  R> v1 <- cf(var = "Y", obs = 0L, sub = c(X = 0L))
  R> v2 <- cf(var = "X", obs = 1L)
  R> v3 <- cf(var = "Z", obs = 0L, sub = c(D = 0L))
  R> v4 <- cf(var = "D", obs = 0L)
  R> c1 <- conj(v1, v2, v3, v4)
  R> g1 <- dag("Y <-> X -> W -> Y <- Z <- D")
  R> g2 <- dag("Y <-> X -> W -> Y <- Z <- D; X -> Y")
  R> out1 <- identifiable(g1, c1)
  R> out2 <- identifiable(g2, c1)
  R> out1
  R> out2
  \end{example}
  \begin{example}
  The query P(y_{x} /\ x'/\ z_{d} /\ d) is identifiable from P_*.
  Formula: \sum_{w} P_{w,z}(y,x')P_{x}(w)P_{d}(z)P(d)
  
  The query P(y_{x} /\ x'/\ z_{d} /\ d) is not identifiable from P_*.
  \end{example}
  The \code{identifiable} function returns an object of class \code{"query"}, whose \code{print} method provides a summary of the identification result. Objects of this class are lists with the following elements: 
  \begin{description}
  \item[\code{id}] A logical value that is \code{TRUE} if the counterfactual query is identifiable and \code{FALSE} otherwise.
  \item[\code{formula}] An object of class \code{"functional"} representing the identifying functional. The \code{format} method for \code{"functional"} objects provides the formula of the counterfactual query in LaTeX syntax when the query is identifiable. Otherwise, \code{formula} is \code{NULL}.
  \item[\code{undefined}] A logical value that is \code{TRUE} if the conditional counterfactual query is found to be undefined.
  \item[\code{query}] The original query as a \code{"counterfactual\_conjunction"} object.
  \item{\code{data}} The \code{data} argument passed to \code{identifiable}.
  \end{description}
  By default, the notation of \citet{shpitser2007} is used for interventional distributions, where interventions are denoted using the subscript, e.g. \(P_x(y)\). If desired, the notation can be swapped to Pearl's notation with the explicit do-operator denoting interventions, e.g., \(P(y|\doo(x))\). This can be accomplished via the \code{use\_do} argument of the \code{format} method for \code{"functional"} objects (passed here via the \code{print} method):
  \begin{example}
  R> print(out1[["formula"]], use_do = TRUE)
  \end{example}
  \begin{example}
  \sum_{w} P(y,x'|do(w,z))P(w|do(x))P(z|do(d))P(d) 
  \end{example}
  For the third example of Section~\ref{sec:id_examples}, we have already defined the counterfactual variable \(Y_x\) of the query as \code{v1} and the observation \(X = x'\) in the condition as \code{v2}. We still need to define the graph of Figure~\ref{fig:graphG} and the other conditioning variable \(Z_x\):
  \begin{example}
  R> g3 <- dag("Z <-> X -> {Z -> Y}")
  R> v5 <- cf("Z", 0, c(X = 0))
  R> identifiable(g3, v1, v5 + v2)
  \end{example}
  \begin{example}
  The query P(y_{x}|z_{x} /\ x') is identifiable from P_*.
  Formula: P_{x,z}(y)
  \end{example}
  Recall from Section~\ref{sec:example}, that this interventional distribution can be further identified, which can be accomplished by setting the \code{data} argument to \code{"observations"} in \code{identifiable} (or to \code{"both"} in this case):
  \begin{example}
  R> identifiable(g3, v1, v5 + v2, data = "observations")
  \end{example}
  \begin{example}
  The query P(y_{x}|z_{x} /\ x') is identifiable from P(v).
  Formula: P(y|x,z)
  \end{example}
  
  \subsection{Formatting output for reports}
  
  The LaTeX formatting of the formulas returned by the \code{identifiable} function enables them to be directly rendered as mathematics, for example in an R Markdown or a Sweave document. For instance, we can write an inline markdown code chunk within a mathematics environment, where we use the \code{format} method with the \code{formula} element of a \code{"query"} object.
  \begin{example}
  \(`r format(out1$formula)`\)
  \end{example}
  This would render as \(\sum_{w} P_{w,z}(y,x')P_{x}(w)P_{d}(z)P(d)\) in the document. Similarly, we could use the \code{use\_do} argument to render the formula such that the do-operator is used instead to represent the interventions.
  \begin{example}
  \(`r format(out1$formula, use_do = TRUE)`\)
  \end{example}
  This would render as \(\sum_{w} P(y,x'|do(w,z))P(w|do(x))P(z|do(d))P(d)\).
  
  Similarly, we can also directly render \code{"counterfactual\_query"} objects into mathematics. We replace the default variable separator string, defined via the \code{var\_sep} argument, with \texttt{" \textbackslash{}\textbackslash{}wedge "} to properly render the \(\wedge\) symbol. The leading and trailing spaces are used so that the command name does not get mixed with the variable names.
  \begin{example}
  \(`r format(c1, var_sep = " \\wedge ")`\)
  \end{example}
  When rendered, this produces \(y_x \wedge x' \wedge z_d \wedge d\).
  
  \section{Summary} \label{sec:summary}
  
  The \pkg{cfid} package provides an easy-to-use interface to identifiability analysis of counterfactual queries. The causal diagram of the causal model can be specified by the user via an intuitive interface, and a variety of commonly used external graph formats are supported. The results from the identifiability algorithms are wrapped neatly in LaTeX syntax to be readily used in publications or reports. This tutorial demonstrates the features of the package and provides insight into the core algorithms it implements.
  
  \section*{Acknowledgments}
  This work was supported by Academy of Finland grant number 331817.
  
  \bibliography{tikka}

\begin{thebibliography}{27}
\providecommand{\natexlab}[1]{#1}
\providecommand{\url}[1]{\texttt{#1}}
\expandafter\ifx\csname urlstyle\endcsname\relax
  \providecommand{\doi}[1]{doi: #1}\else
  \providecommand{\doi}{doi: \begingroup \urlstyle{rm}\Url}\fi

\bibitem[Avin et~al.(2005)Avin, Shpitser, and Pearl]{avin2005pathspecific}
C.~Avin, I.~Shpitser, and J.~Pearl.
\newblock Identifiability of path-specific effects.
\newblock In \emph{Proceedings of International Joint Conference on Artificial
  Intelligence}, volume~19, pages 357--363, 01 2005.

\bibitem[Balke and Pearl(1994{\natexlab{a}})]{balke1994a}
A.~Balke and J.~Pearl.
\newblock Counterfactual probabilities: Computational methods, bounds and
  applications.
\newblock In \emph{Proceedings of the 10th Conference on Uncertainty in
  Artificial Intelligence}, pages 46--54, 1994{\natexlab{a}}.

\bibitem[Balke and Pearl(1994{\natexlab{b}})]{balke1994b}
A.~Balke and J.~Pearl.
\newblock Probabilistic evaluation of counterfactual queries.
\newblock In \emph{Proceedings of the 12th AAAI National Conference on
  Artificial Intelligence}, pages 230--237, 1994{\natexlab{b}}.

\bibitem[Bareinboim and Pearl(2012)]{bareinboim2012zid}
E.~Bareinboim and J.~Pearl.
\newblock Causal inference by surrogate experiments: $z$-identifiability.
\newblock In \emph{Proceedings of the 28th Conference on Uncertainty in
  Artificial Intelligence}, pages 113--120, 2012.

\bibitem[Chen et~al.(2020)Chen, Chernozhukov, Fernandez-Val, and
  Melly]{counterfactualpackage}
M.~Chen, V.~Chernozhukov, I.~Fernandez-Val, and B.~Melly.
\newblock \emph{{Counterfactual}: Estimation and Inference Methods for
  Counterfactual Analysis}, 2020.
\newblock URL \url{https://CRAN.R-project.org/package=Counterfactual}.
\newblock {R}~package version~1.2.

\bibitem[Csardi and Nepusz(2006)]{igraph}
G.~Csardi and T.~Nepusz.
\newblock The {igraph} software package for complex network research.
\newblock \emph{InterJournal}, Complex Systems:\penalty0 1695, 2006.
\newblock URL \url{https://igraph.org}.

\bibitem[Halpern(1998)]{halpern1998}
J.~Y. Halpern.
\newblock Axiomatizing causal reasoning.
\newblock In \emph{Proceedings of the 14th Conference on Uncertainty in
  Artificial Intelligence}, pages 202--210, 1998.

\bibitem[Holland(1986)]{holland1986}
P.~W. Holland.
\newblock Statistics and causal inference.
\newblock \emph{Journal of the American Statistical Association}, 81\penalty0
  (396):\penalty0 945--960, 1986.
\newblock URL \url{https://doi.org/10.1080/01621459.1986.10478354}.

\bibitem[Huang and Valtorta(2006)]{huang2006complete}
Y.~Huang and M.~Valtorta.
\newblock Pearl's calculus of intervention is complete.
\newblock In \emph{Proceedings of the 22nd Conference on Uncertainty in
  Artificial Intelligence}, pages 217--224. AUAI Press, 2006.

\bibitem[Karvanen(2022)]{r6causal}
J.~Karvanen.
\newblock \emph{{R6causal}: R6 Class for Structural Causal Models}, 2022.
\newblock {R}~package version~0.6.1.

\bibitem[Kivva et~al.(2022)Kivva, Mokhtarian, Etesami, and Kiyavash]{kivva2022}
Y.~Kivva, E.~Mokhtarian, J.~Etesami, and N.~Kiyavash.
\newblock Revisiting the general identifiability problem.
\newblock In \emph{Proceedings of the 38th Conference on Uncertainty in
  Artificial Intelligence}, volume 180, pages 1022--1030. PMLR, 2022.

\bibitem[Kusner et~al.(2017)Kusner, Loftus, Russell, and
  Silva]{KusnerCounterfactual}
M.~J. Kusner, J.~Loftus, C.~Russell, and R.~Silva.
\newblock Counterfactual fairness.
\newblock In \emph{Proceedings of the 31st International Conference on Neural
  Information Processing Systems}, pages 4069--4079, 2017.

\bibitem[Lee et~al.(2019)Lee, Correa, and Bareinboim]{lee2019surrogate}
S.~Lee, J.~D. Correa, and E.~Bareinboim.
\newblock General identifiability with arbitrary surrogate experiments.
\newblock In \emph{Proceedings of the 35th Conference on Uncertainty in
  Artificial Intelligence}, volume 115, pages 389--398. PMLR, 2019.

\bibitem[Pearl(1995)]{pearl1995}
J.~Pearl.
\newblock Causal diagrams for empirical research.
\newblock \emph{Biometrika}, pages 669--710, 1995.
\newblock URL \url{https://doi.org/10.1093/biomet/82.4.669}.

\bibitem[Pearl(2009)]{pearl2009}
J.~Pearl.
\newblock \emph{Causality: Models, Reasoning and Inference}.
\newblock Cambridge University Press, 2nd edition, 2009.

\bibitem[Shpitser and Pearl(2006{\natexlab{a}})]{shpitser2006id}
I.~Shpitser and J.~Pearl.
\newblock Identification of joint interventional distributions in recursive
  semi-{M}arkovian causal models.
\newblock In \emph{Proceedings of the 21st National Conference on Artificial
  Intelligence - Volume 2}, pages 1219--1226. AAAI Press, 2006{\natexlab{a}}.

\bibitem[Shpitser and Pearl(2006{\natexlab{b}})]{shpitser2006idc}
I.~Shpitser and J.~Pearl.
\newblock Identification of conditional interventional distributions.
\newblock In \emph{Proceedings of the 22nd Conference on Uncertainty in
  Artificial Intelligence}, pages 437--444. AUAI Press, 2006{\natexlab{b}}.

\bibitem[Shpitser and Pearl(2007)]{shpitser2007}
I.~Shpitser and J.~Pearl.
\newblock What counterfactuals can be tested.
\newblock In \emph{Proceedings of the 23rd Conference on Uncertainty in
  Artificial Intelligence}, pages 352--359. AUAI Press, 2007.

\bibitem[Shpitser and Pearl(2008)]{shpitser2008}
I.~Shpitser and J.~Pearl.
\newblock Complete identification methods for the causal hierarchy.
\newblock \emph{Journal of Machine Learning Research}, 9\penalty0
  (64):\penalty0 1941--1979, 2008.

\bibitem[Stoll et~al.(2020)Stoll, King, Zeng, Gandrud, and
  Sabath]{whatifpackage}
H.~Stoll, G.~King, L.~Zeng, C.~Gandrud, and B.~Sabath.
\newblock \emph{{WhatIf}: Software for Evaluating Counterfactuals}, 2020.
\newblock URL \url{https://CRAN.R-project.org/package=WhatIf}.
\newblock {R}~package version~1.5-10.

\bibitem[Textor et~al.(2017)Textor, van~der Zander, Gilthorpe, Li\'{s}kiewicz,
  and Ellison]{dagitty}
J.~Textor, B.~van~der Zander, M.~S. Gilthorpe, M.~Li\'{s}kiewicz, and G.~T.
  Ellison.
\newblock Robust causal inference using directed acyclic graphs: The {R}
  package {dagitty}.
\newblock \emph{International Journal of Epidemiology}, 45\penalty0
  (6):\penalty0 1887--1894, 2017.
\newblock URL \url{https://doi.org/10.1093/ije/dyw341}.

\bibitem[Tian and Pearl(2002)]{tian2002general}
J.~Tian and J.~Pearl.
\newblock A general identification condition for causal effects.
\newblock In \emph{Proceedings of the 19th AAAI National Conference on
  Artificial Intelligence}, pages 567--573, 2002.

\bibitem[Tikka and Karvanen(2017)]{tikka2017}
S.~Tikka and J.~Karvanen.
\newblock Identifying causal effects with the {R} package {causaleffect}.
\newblock \emph{Journal of Statistical Software}, 76\penalty0 (12):\penalty0
  1--30, 2017.
\newblock URL \url{https://doi.org/10.18637/jss.v076.i12}.

\bibitem[Tikka and Karvanen(2019)]{tikka2019surrogate}
S.~Tikka and J.~Karvanen.
\newblock Surrogate outcomes and transportability.
\newblock \emph{International Journal of Approximate Reasoning}, 108:\penalty0
  21--37, 2019.

\bibitem[Tikka et~al.(2021)Tikka, Hyttinen, and Karvanen]{dosearch}
S.~Tikka, A.~Hyttinen, and J.~Karvanen.
\newblock Causal effect identification from multiple incomplete data sources: A
  general search-based approach.
\newblock \emph{Journal of Statistical Software}, 99\penalty0 (5):\penalty0
  1--40, 2021.
\newblock URL \url{https://doi.org/10.18637/jss.v099.i05}.

\bibitem[Verma and Pearl(1990)]{verma1990}
T.~S. Verma and J.~Pearl.
\newblock Equivalence and synthesis of causal models.
\newblock In \emph{Proceedings of the 6th Conference on Uncertainty in
  Artificial Intelligence}, pages 255--270, 1990.

\bibitem[Zhang and Bareinboim(2018)]{ZhangBareinboim2018}
J.~Zhang and E.~Bareinboim.
\newblock Fairness in decision-making --- the causal explanation formula.
\newblock In \emph{Proceedings of the 32nd AAAI Conference on Artificial
  Intelligence}, pages 2037--2045, 2018.

\end{thebibliography}
  
  \address{Santtu Tikka\\
  Department of Mathematics and Statistics, University of Jyv\"askyl\"a\\
  P.O. Box 35, FI-40014, Finland\\
  ORCiD: \href{https://orcid.org/0000-0003-4039-4342}{0000-0003-4039-4342}\\
  \email{santtu.tikka@jyu.fi}}  

\end{article}

\end{document}